\documentclass[%
reprint,
 amsmath,amssymb,
 aps,
]{revtex4-2}

\usepackage{graphicx}% Include figure files
\usepackage{dcolumn}% Align table columns on decimal point
\usepackage{bm}% bold math
\usepackage{comment}
\usepackage{xcolor}

\usepackage[a4paper, total={7.3in, 10.65in}]{geometry}
\usepackage{hyperref}% add hypertext capabilities
\hypersetup{unicode=true,bookmarksnumbered=true,bookmarksopen=true,
 breaklinks=false,pdfborder={0 0 0},colorlinks=true,
 linkcolor=blue, citecolor=blue, urlcolor=blue}

\begin{document}

\title{Ultrafast TACOS -- Terahertz-Assisted Chiro-Optical Spectroscopy}% 

\author{J. Terentjevas$^{1,2}$, P. Vindel-Zandbergen$^{3}$, L. Rego$^{2,4,5}$, F. Morales$^{1}$, A. Ord\'o\~nez$^{2,6}$, O. Smirnova$^{1,7}$, D. Ayuso$^{1,2,6}$}
\affiliation{
 $^{1}$Max-Born-Institut, 12489 Berlin, Germany}
\affiliation{ $^{2}$Department of Physics, Imperial College London, SW7 2BW London, UK
}%
\affiliation{$^{3}$Department of Chemistry, New York University, New York 10003, New York, USA}
\affiliation{$^{4}$Instituto Madrileño de Estudios Avanzados en Nanociencia (IMDEA Nano), Cantoblanco 28049, Madrid, Spain}
\affiliation{$^{5}$Departamento de Química, Universidad Autónoma de Madrid, 28049 Madrid, Spain}
\affiliation{$^{6}$School of Physical and Chemical Sciences, Queen Mary University of London, E1 4NS, UK}
\affiliation{$^{7}$Technische Universität Berlin, 10623 Berlin, Germany}

\begin{abstract}
We bring together the advantages of terahertz (THz) and optical spectroscopies to introduce TACOS (Terahertz-Assisted Chiro-Optical Spectroscopy), a novel approach for ultrafast and highly efficient imaging of molecular chirality and control over chiral electronic dynamics. We show how, using a THz pulse, we can induce a transient electronic orientation in a medium of randomly oriented chiral molecules that breaks the isotropy of the molecular sample. This symmetry breaking twists the nonlinear response of the medium to an ultrashort linearly polarised optical pulse in a highly enantiosensitive manner. As a result, the medium emits elliptically polarised light at new optical frequencies that records the molecular handedness via purely electric-dipole interactions. The long wavelength and period of the THz pulse enable both spatial coherence across the sample and a substantial degree of electronic orientation over the duration of the ultrashort optical pulse. TACOS does not require optical carrier-envelope phase stability or working in vacuum, and it creates exciting avenues for ultrafast and highly efficient chiral sensing and manipulation.
\end{abstract}

\maketitle

Opposite versions of a chiral object display non-superimposable mirror images of each other.
This geometrical property makes them behave identically, except when interacting with another chiral entity.
For instance, whereas a sock can fit both feet, only the right shoe fits the right foot.
Chirality is ubiquitous in biological systems, and thus telling apart opposite versions of the same chiral molecule (enantiomers) is vital for pharmaceutical developments and even for medical diagnosis \cite{liu_detection_2023}, but also challenging.
Traditional optical methods \cite{berova_application_2007,schellman_circular_1975, nafie_vibrational_1976,  Costante_Raman_1997, polavarapu_absolute_1998, orlova_polarimetry_2014,tran_continuous-wave_2021}, such as polarimetry \cite{orlova_polarimetry_2014,tran_continuous-wave_2021} and photo-absorption  circular dichroism \cite{berova_application_2007, schellman_circular_1975} 
rely on the weak magnetic coupling between light and matter, which leads to weak enantiosensitivity \cite{Barron2004} and creates important challenges, particularly for ultrafast spectroscopy. 

The electric-dipole ``revolution'' \cite{ayuso_ultrafast_2022} is transforming the landscape of molecular chiral discrimination with a new generation of chiro-optical methods offering orders-of-magnitude greater enantiosensitivity.
Different ``streams'' of this revolution rely on analysing different enantiosensitive observables, from microwave radiation \cite{patterson_enantiomer-specific_2013,Patterson_sensitive_2013, eibenberger_enantiomer-specific_2017,perez_coherent_2017, leibscher_principles_2019} to optical/UV signals \cite{fischer_chiral_2002,ayuso_synthetic_2019, ayuso_enantio-sensitive_2021,ayuso_ultrafast_2021,ayuso_new_2022,ayuso_strong_2022,khokhlova_enantiosensitive_2022, rego_tilting_2023} or to photoelectron currents \cite{ritchie_theory_1976,powis_photoelectron_2000,bowering_asymmetry_2001,garcia_circular_2003,garcia_vibrationally_2013,janssen_detecting_2014,lux_circular_2012,lehmann_imaging_2013,lux_photoelectron_2015,kastner_enantiomeric_2016,comby_relaxation_2016,beaulieu_probing_2016,comby_real-time_2018,demekhin_photoelectron_2018,
goetz_quantum_2019,rozen_controlling_2019,ordonez_disentangling_2022, beaulieu_multiphoton_2018,beaulieu_photoexcitation_2018,Planas_vortices_2022, wanie_dynamics_2024}, but share one common ingredient \cite{ordonez_generalized_2018}: the enantiosensitive response of the molecules is driven solely by the electronic response of the molecules to the local polarisation of the electric driving field -- magnetic interactions are not required.

Locally chiral light \cite{ayuso_synthetic_2019, ayuso_enantio-sensitive_2021,ayuso_ultrafast_2021,ayuso_new_2022,ayuso_strong_2022}, where the electric field draws a chiral 3D Lissajous figure in time, can drive enantiosensitive electronic dynamics via purely electric-dipole interactions. 
This type of light can be created by overlapping two laser beams that carry several frequencies and propagate non-collinearly.
The non-collinear nature of the setup creates a spatial modulation of the overall field's properties, with a periodicity on the order of the laser wavelengths \cite{ordonezChiralCoherentControl2023}.
This can be exploited to steer the light emission from the chiral molecules in an enantiosensitive way \cite{ayuso_enantio-sensitive_2021,Rego2023NJP}.
However, since such periodicities are on the order of the optical wavelengths, and thus much smaller than typical interaction regions, one needs to carefully design experiments so that such spatial modulations do not wash out enantiosensitive effects in the macroscopic response of the medium.

The microwave counterpart of locally chiral light can also be created in non-collinear configurations, but it maintains its properties over spatial regions that are on the order of microwave wavelengths, which can cover the molecular samples.
Locally chiral microwave radiation can drive highly enantiosensitive \emph{rotational} dynamics in chiral molecules \cite{eibenberger_enantiomer-specific_2017, lee_quantitative_2022}. However, microwaves are not well suited for attochemistry, an emerging field which aims to image and control chemical reactions by manipulating \emph{electronic} motion in molecules on the attosecond timescale \cite{Calegari_attochemistry_2023}.

\begin{figure*}[t]
\includegraphics[width=\textwidth]{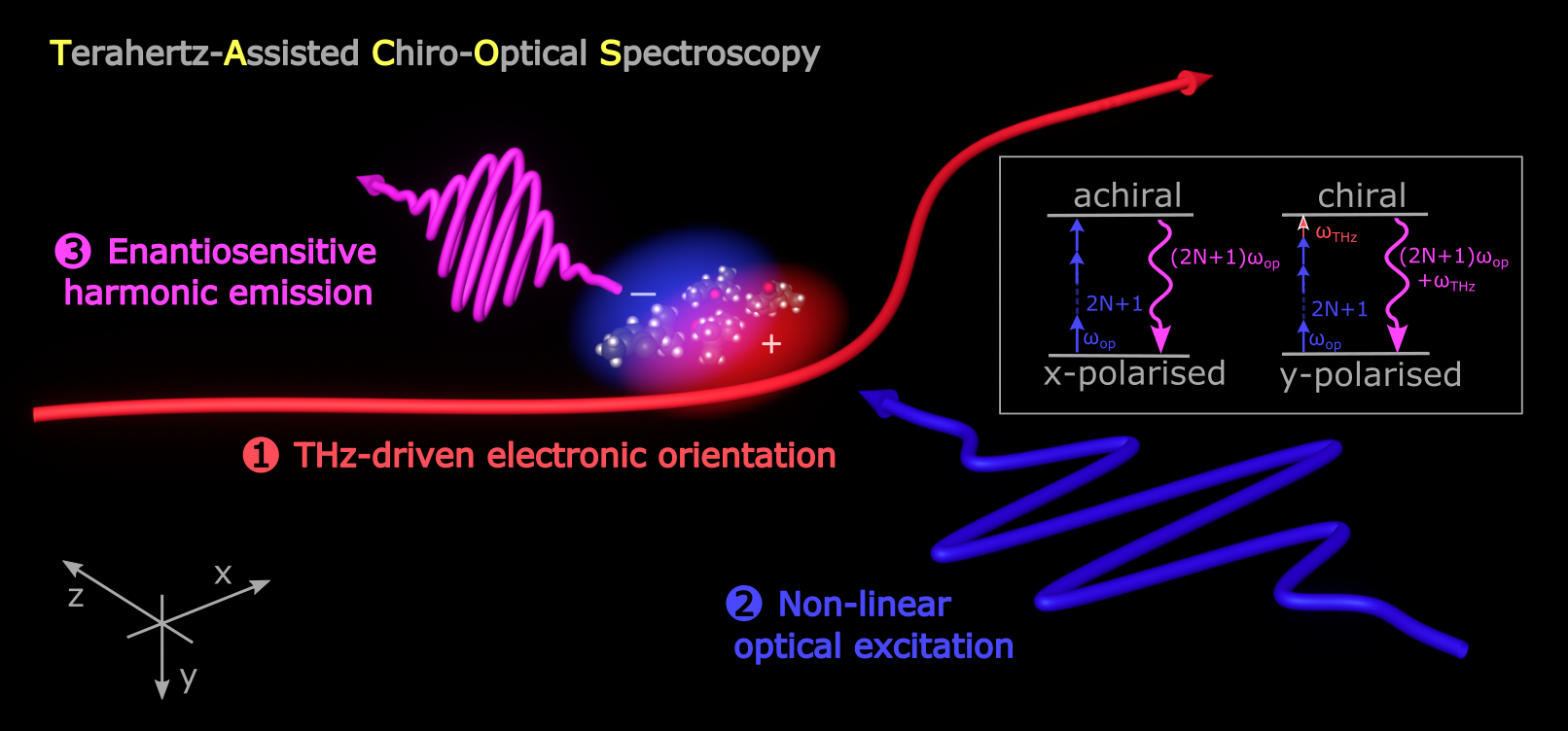}
\caption{\label{fig_picture} 
\textbf{Physical picture. Step 1:}
a linearly-polarized THz field (red) transiently orients the electronic density in a randomly oriented ensemble of chiral molecules via linear response, making the medium anisotropic.
\textbf{Step 2}: the THz-driven electronic orientation twists the non-linear response of the medium to an ultrashort cross-polarised and cross-propagating optical pulse (blue).
\textbf{Step 3}: the radiation emitted by the medium at new optical frequencies (magenta) acquires enantiosensitive elliptical polarisation: the  ellipticity and rotation (tilt) angle record the molecular handedness via purely electric-dipole interactions.
The relevant achiral and chiral multiphoton pathways are depicted in the inset.}
\end{figure*}

Can we bring the advantages of both approaches together?
That is, can we use optical frequencies to drive ultrafast electronic dynamics in chiral molecules, while benefiting from the long wavelengths of microwave or terahertz radiation?
We provide a positive answer to this question by introducing terahertz-assisted chiro-optical spectroscopy (TACOS).

Our approach involves cross-polarised and cross-propagating THz and optical pulses interacting with a medium of randomly oriented chiral molecules, see Fig 1.
Without the THz field, the local response of the medium to the optical pulse would be linearly polarised and insensitive to its handedness.
The linear response to the THz field breaks this symmetry by transiently orienting the electronic clouds of the molecules. As a result, the non-linear response to the optical pulse acquires an enantiosensitive polarisation component that is orthogonal to both driving fields and enables efficient chiral recognition: it leads to emission of harmonic light which has opposite polarisation in opposite enantiomers.

Using the linear response to a THz pulse to electronically orient the molecules ensures that they will maintain the same orientation over a spatial region on the order of half the THz wavelength and that the orientation will persist during half the THz period. 
Thus, if either the optical beam at the focus or the medium, or a substantial part of them, are smaller than half the THz wavelength, the enantiosensitive electronic response of the medium can be maintained in space over the whole interaction region.
TACOS also requires few-cycle optical pulses, although optical carrier-envelope phase (CEP) stability is not needed and the THz field does not need to be ultrashort. 

Let us consider a multiphoton picture and focus on low-order non-linear processes resulting from purely electric-dipole interactions.
The molecules can absorb 3 photons from the $x$-polarised optical pulse, see Fig. 1 (inset), leading to achiral polarisation at frequency $3\omega_{\text{op}}$ along $x$.
Since we are neglecting magnetic interactions, this pathway is not sensitive to the medium's handedness \cite{fischer_nonlinear_2005}.
The next-order non-linear process involving 3 optical photons and 1 additional $z$-polarised THz photon induces polarisation at frequency $3\omega_{\text{op}}+\omega_{\text{THz}}$ along $y$.
This even-order response is exclusive of chiral media and it has an opposite phase in opposite enantiomers.
It can be understood as chiral sum-frequency generation \cite{Giordmaine_SFG_1965, fischer_three-wave_2000, belkin_sum-frequency_2000, fischer_chiral_2002, fischer_nonlinear_2005, Vogwell_SFG_2023} from the optical and THz fields assisted by the absorption of two additional optical photons.
Using a few-cycle optical pulse (a key ingredient of TACOS) with a spectral bandwidth $\Delta\omega_{\text{op}}>\omega_{\text{THz}}$, enables the interference between the achiral and chiral pathways in the frequency domain.
As a result, the non-linear response of the medium around the frequency of $3\omega_{\text{op}}$ acquires an enantiosensitive polarisation.
Higher order nonlinear processes involving additional $2N$ optical photons in both pathways can also be enantiosensitive.

Having $\omega_{\text{op}}>>\omega_{\text{THz}}$ not only facilitates the spectral overlap between the achiral and chiral pathways (Fig. \ref{fig_picture}) in the near field, it also ensures an effective far-field interference.
Indeed, $|\textbf{k}_{\text{op}}|>>|\textbf{k}_{\text{THz}}|$ implies that $3\textbf{k}+\textbf{k}_{\text{THz}}\simeq3\textbf{k}$, and thus the two pathways lead to emission along essentially the same direction.
As a result, the emitted radiation maintains its enantiosensitive polarisation as it reaches the detector.

We have numerically tested our proposal in randomly oriented carvone molecules, see Methods, considering the following optical parameters: wavelength $\lambda_{\mathrm{op}} = 660$nm, intensity $I_{\mathrm{op}} = 10^{12}$W/cm$^2$ and pulse duration $\tau_{\mathrm{op}}=10$fs (FWHM).
For the THz field, we used $\lambda_{\mathrm{THz}} = 80\mu$m (period $T_{\mathrm{THz}}=267$fs),  $I_{\mathrm{THz}} = 10^{12}$W/cm$^2$ and $\tau_{\mathrm{THz}}=355$fs.
We used a relatively short wavelength and an ultrashort duration in the THz pulse to keep the simulations affordable, but we note this is not a requirement.
Both parameters can be increased without degrading the efficiency of the chiral discrimination.
The wavelength of the optical pulse was chosen to be in electronic resonance in order to maximise the enantiosensitive response.

\begin{figure}[h!]
\includegraphics[width=0.45\textwidth]{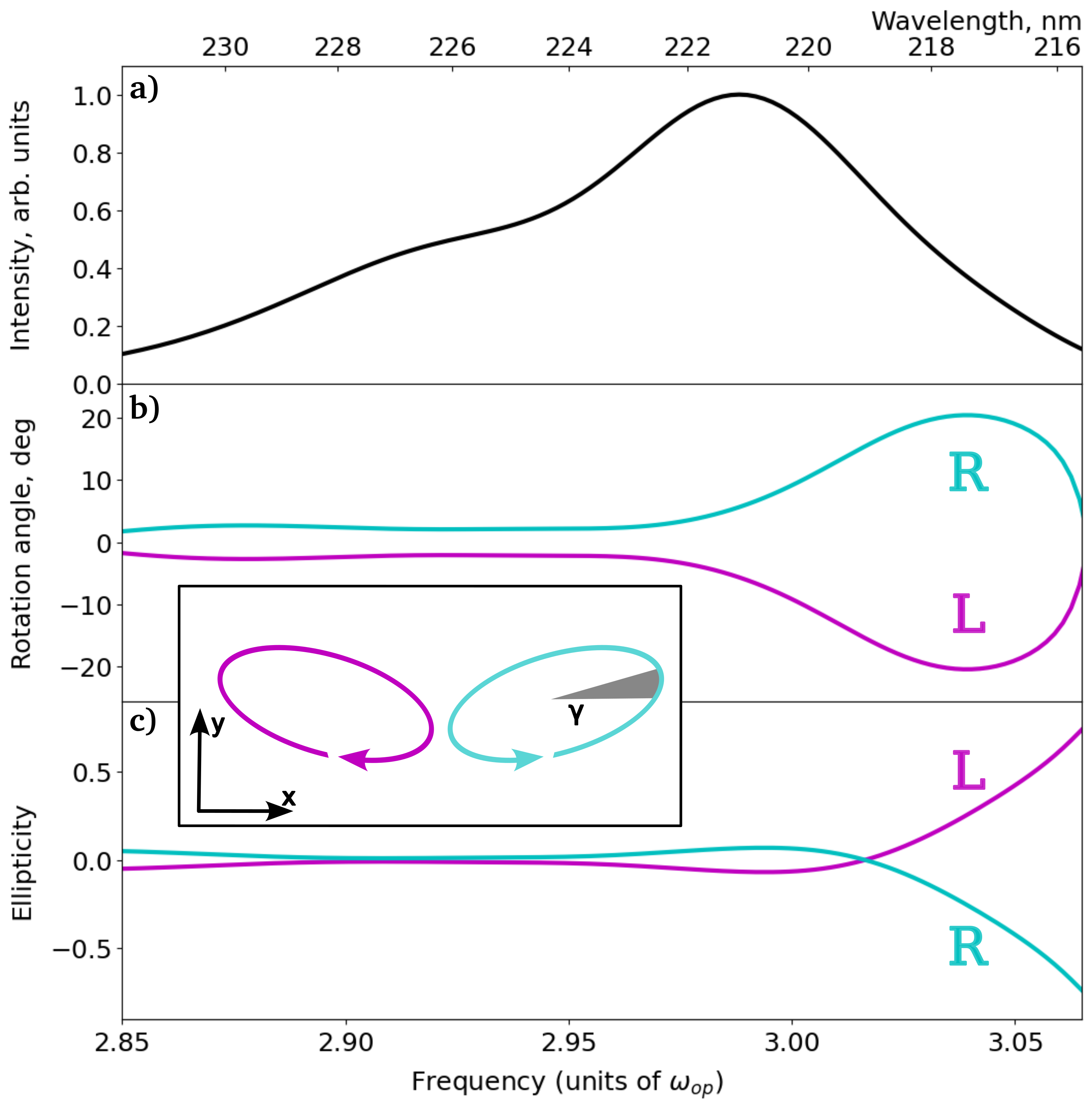}
\caption{\label{fig_results1D}
\textbf{Non-linear response of randomly oriented carvone ($\boldsymbol{\mathrm{CEP} = 0}$)}.
\textbf{a} Intensity of light emission in the vicinity of $3\omega_{\mathrm{op}}$ as a function of the output frequency (lower axis) and wavelength (upper axis) driven by the setup of Fig. \ref{fig_picture}.
The intensity of emission is not enantiosensitive because the overall driving field is achiral.
\textbf{b,c,} Enantiosensitive polarisation of the radiation emitted from the L (magenta) and R (cyan) enantiomers: rotation angle of the major component of the polarisation ellipse (\textbf{b}) and ellipticity (\textbf{c}). Here ellipticity is defined as $b/a$, where $a$ and $b$ are the semimajor and semiminor axes respectively.
}

\end{figure}

\begin{figure*}
\includegraphics[width=0.89\textwidth]{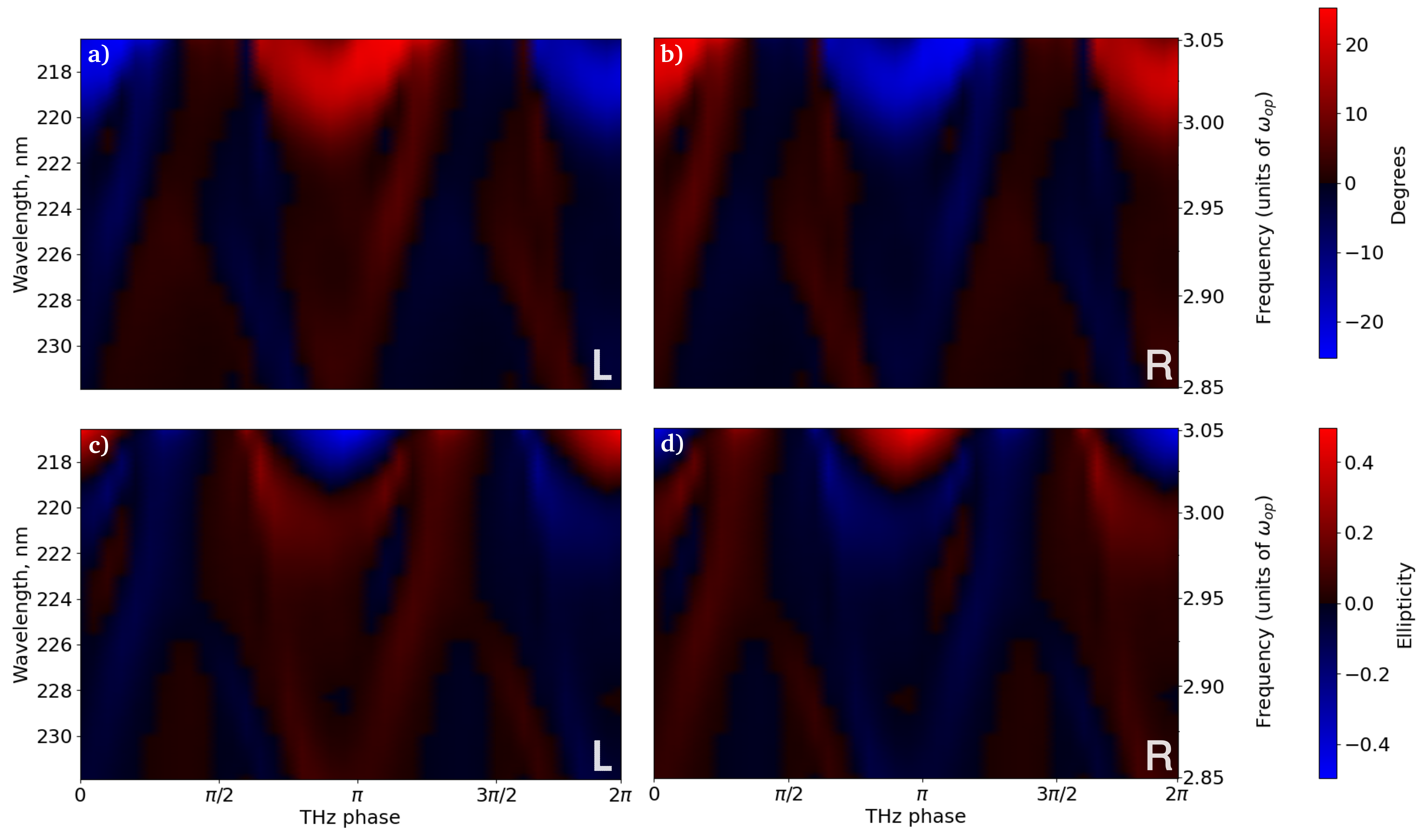}
\caption{\label{fig_results2D}
\textbf{Enantiosensitive control over the polarisation of the nonlinear optical response}.
Rotation angle of the major component of the polarisation ellipse (\textbf{a}, \textbf{b}) and ellipticity (\textbf{c}, \textbf{d}) of the radiation emitted from the L (\textbf{a}, \textbf{c}) and R (\textbf{b}, \textbf{d}) enantiomers of carvone as a function of the output wavelength/frequency and the phase of the THz field. In our calculations, we use an ultrashort THz pulse and vary its CEP, but using few-cycle THz pulses is not a requirement. If TACOS is implemented using a longer THz field, a sub-cycle temporal delay would lead to the same effect.
}
\end{figure*}

Fig. \ref{fig_results1D} shows the intensity and polarisation of the light emitted from the two enantiomers of carvone in the vicinity of $3\omega_{\mathrm{op}}$, where the two quantum pathways (Fig. \ref{fig_picture}) effectively interfere.
The intensity (Fig. \ref{fig_results1D}a) is not enantiosensitive because the overall driving field is achiral: it has mirror symmetry with respect to the $xy$ plane.
Yet, it can encode the molecular handedness in the polarisation of the emitted light, see Fig. \ref{fig_results1D}b,c.
Indeed, because the x-component of the emitted radiation is not sensitive to the medium's handedness and the y-component has opposite phase in opposite enantiomers, the ellipticity $\varepsilon$ (Fig. \ref{fig_results1D}c) and the rotation, or tilt, angle $\gamma$ of the polarisation ellipse (Fig. \ref{fig_results1D}b) record the chirality of the medium.
Notably, the relative rotation angle $\gamma_R-\gamma_L$ reaches $15^\circ$ degrees at the emission peak, and it becomes larger at higher frequencies.
We also note that the emitted light at wavelength $\lambda \approx 220$nm is not substantially absorbed by air, which means that our proposal does not require working in vacuum.

We now show how we can control the enantiosensitive response of the medium by controlling the properties of the incident fields.
The polarisation of the emitted radiation records the relative amplitude and phase between the induced polarisation associated with the achiral and chiral pathways (Fig. \ref{fig_picture}, inset).
Because these pathways involve absorption of the same number of optical photons, their relative amplitude and phase are not sensitive to the intensity and phase of the optical pulse.
This means that the enantiosensitive response is not sensitive to the CEP of the optical pulse, and thus optical CEP stability is not needed.
However, because only the chiral pathway is sensitive to the THz field (it involves the absorption of one THz photon), we can control the enantiosensitive response by controlling the properties of the THz pulse.
Indeed, changing the THz phase changes the phase of the induced polarisation associated with the chiral pathway without altering the amplitude of the chiral response.

Fig. \ref{fig_results2D} shows the rotation angle, or tilt, of the polarisation ellipse (Fig. \ref{fig_results2D}a,b) and ellipticity (Fig. \ref{fig_results2D}c,d) of the emitted radiation as functions of the THz CEP and output frequency in the vicinity of $3\omega_{\mathrm{op}}$, for (L)- and (R)-carvone.
We find large tilt angles and ellipticities across the spectrum, reaching up to $\gamma\simeq \pm22^{\circ}$ and $\varepsilon\simeq \pm0.5$.
Note that changing the molecular enantiomer changes the phase associated with the chiral pathway without altering the phase of the achiral pathway, and thus both $\gamma$ and $\varepsilon$ flip sign when exchanging the molecular enantiomers.
Changing the THz CEP by $\pi$ leads to the same effect.

\begin{figure}
\includegraphics[width=0.48\textwidth]{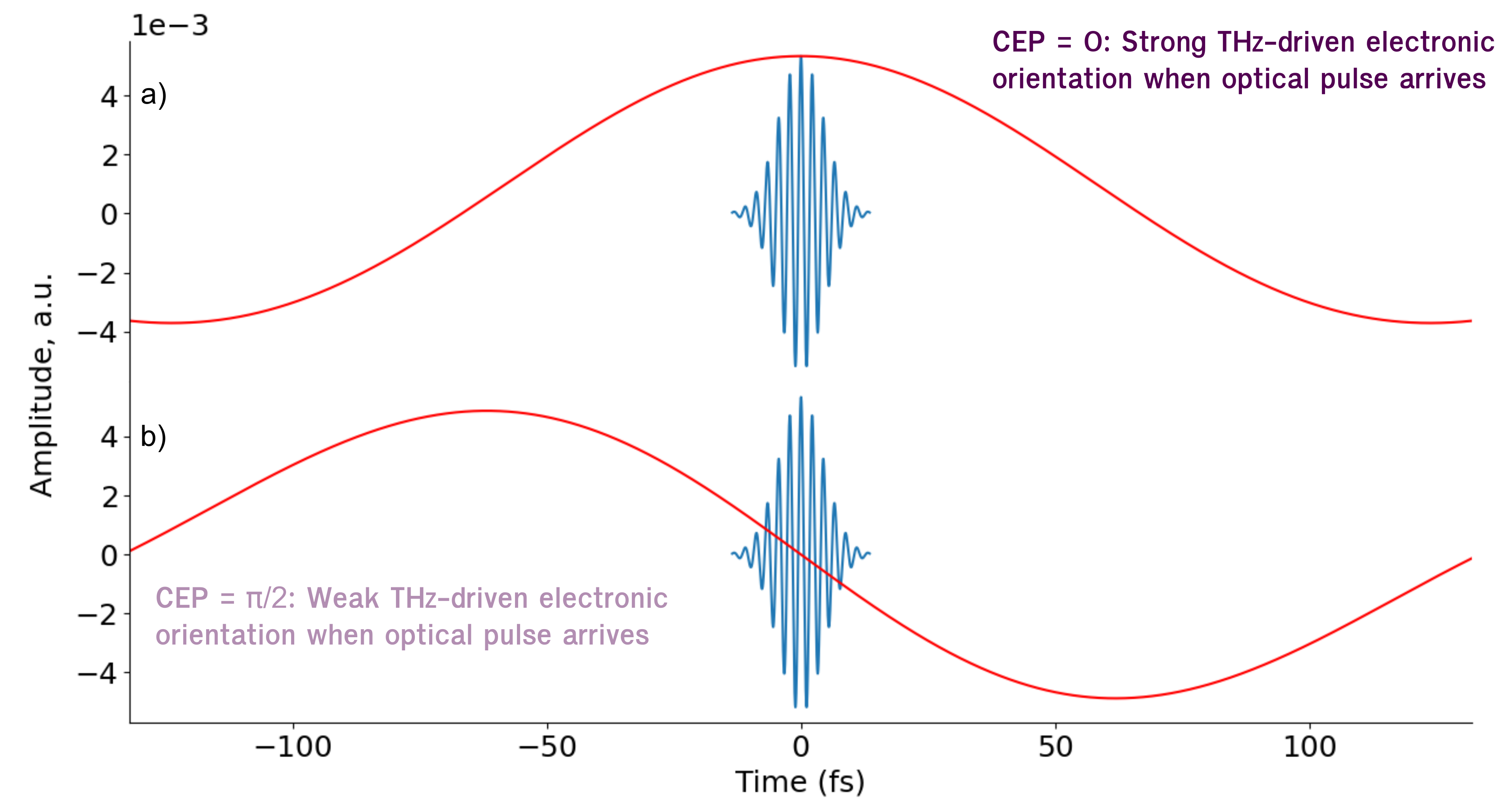}
\caption{\label{fig_cep}
\textbf{Understanding the generation of the chiro-optical response.}
Electric-field amplitude of the optical (blue) and THz (red) fields when the CEP of the THz field is 0 (\textbf{a}) and $\pi/2$ (\textbf{b}).
The linear response to the THz field induces transient electronic orientation in the medium.
\textbf{a,} If CEP=0 (or $\pi$), the ultrashort optical pulse arrives to the medium when the degree of THz-driven electronic orientation is at its maximum, resulting in a strong chiro-optical signal (see Fig. \ref{fig_results2D}).
\textbf{b,} If CEP=$\pi/2$ (or $3\pi/2$) the optical field arrives to the sample when it is barely oriented, hence producing a weak chiro-optical response.
}
\end{figure}

The modulation of the enantiosensitive response with the THz phase records the transient electronic orientation driven by the THz field in the medium.
In our calculations, we assume that the temporal envelopes of the THz and optical pulses are centered.
Thus, if the THz CEP is either $0$ or $\pi$, the maximum of the optical pulse envelope coincides with a maximum or minimum of the THz field amplitude, see Fig. \ref{fig_cep}a.
In these situations, the optical pulse arrives to the sample when the degree of THz-driven electronic orientation is at the maximum, leading to a strong chiro-optical response.
However, if the THz CEP is either $0.5\pi$ or $1.5\pi$, the maximum of the optical pulse envelope coincides with a zero in the THz field amplitude, see Fig. \ref{fig_cep}b.
That is, the optical pulse arrives when the degree of electronic orientation is weak, and thus it produces a weak chiro-optical signal. This can be seen in Fig. \ref{fig_results2D}.

We emphasise that, while we use ultrashort THz pulses to keep our simulations affordable, this is not a requirement of TACOS.
We can obtain equivalent results using a longer THz pulse and varying the temporal delay between THz and optical fields, as long as the THz pulse is sufficiently intense to drive a substantial degree of transient electronic orientation in the medium.

Recent developments in the generation of intense THz pulses \cite{hoffmann_intense_2011,koulouklidis_observation_2020,kim_strong_2016} from ultrashort optical pulses \cite{kim_strong_2016,Jang_20,Nam_23} may facilitate the experimental realisation of TACOS.
Note that the THz field plays the role of a quasi-static electric field, which induces an electronic orientation of the chiral molecules.
The THz period is sufficiently long so the molecules remain electronically oriented during the whole interaction with the optical pulse, but also sufficiently short so the molecules do not undergo substantial ionisation -- as it would inevitably happen if one attempts to realise our proposal using a constant electric field of comparable amplitude.

TACOS combines short- and long-wavelength radiation to drive highly enantiosensitive electronic dynamics in chiral molecules on ultrafast timescales, bringing together the advantages of THz and optical spectroscopies.
The long wavelength and period of the THz field enable both a large spatial coherence across the sample and a transient electronic orientation that survives the duration of the optical pulse.
Further implementations of TACOS using tailored polarisation either in the optical or THz field may lead to richer chiral dynamics, stronger chiro-optical signals, and higher degrees of enantiosensitive control in attochemistry.

\section*{Acknowledgements}

We acknowledge enlightening discussions with Misha Ivanov.
J.T., A.O. and D.A. acknowledge funding from the Royal Society URF$\backslash$R1$\backslash$201333, URF$\backslash$ERE$\backslash$210358 and URF$\backslash$ERE$\backslash$231177.
L. R. acknowledges financial support from the European Union-NextGenerationEU and the Spanish Ministry of Universities via Margarita Salas Fellowship through the University of Salamanca.
J.T. and O.S. acknowledge funding from the European Union (ERC, ULISSES, 101054696). J.T. and D.A. acknowledge funding from COST Action CA18222, supported by COST (European Cooperation in Science
and Technology). Views and opinions expressed are, however, those of the authors only and do not necessarily reflect those of the European Union or the European Research Council. Neither the European Union nor the granting authority can be held responsible for them.

\section*{Methods}

The molecular geometry of the most stable conformer of (R)-carvone was taken from Ref. \citenum{lambert_optical_2012} and reoptimized for consistency using Density Functional Theory (DFT) \cite{kohn_density_1996,ziegler_approximate_2002} with
the B3LYP \cite{lee_development_1988,becke_density-functional_1988}
functional and the 6-311G(d,p) \cite{mclean_contracted_2008} basis
set available in Gaussian 16 \cite{g16}.
The excitation energies and transition dipoles were evaluated for the first 100 excited states using time-dependent DFT at the CAM-B3LYP/d-aug-cc-pVDZ level of theory \cite{CAMB3LYP,CAMB3LYP_2,augccpvdz_2,augccpvdz_3}.
The transition dipoles between the excited states were computed with the Multiwfn software \cite{Multiwfn}.

The ultrafast electronic response of the most stable conformer of carvone was evaluated by solving numerically the time-dependent Schr\"{o}dinger equation (TDSE) in the basis set of the field-free eigenstates using an explicit Runge-Kutta method of order 8 in the presence of the THz and optical fields:
\begin{equation}
    \boldsymbol{E} = E_{\mathrm{op}}(t)\cos(\omega_{\mathrm{op}}t)\hat{\boldsymbol{x}} + E_{\mathrm{THz}}(t)\cos(\omega_{\mathrm{THz}}t+\Delta)\hat{\boldsymbol{z}}
\end{equation}
where $E_{\mathrm{op}}$ and $E_{\mathrm{THz}}$ are the Gaussian envelopes for the optical ($\sigma = 182$ a.u.) and terahertz ($\sigma = 6230$ a.u.) pulses, respectively, each of peak amplitude $E_{max} = 0.00534$ a.u. $\omega_{\mathrm{op}} = 0.0690$ a.u. and $\omega_{\mathrm{THz}} = 0.000570$ a.u. are the optical and terahertz pulse frequencies, respectively, and $\Delta$ is the CEP of the terahertz pulse.
To describe a randomly orientated ensemble, we ran TDSE simulations for 208 molecular orientations and averaged the induced polarisation
\begin{equation}
    \boldsymbol{P} = \frac{1}{8\pi^{2}}\int_{0}^{2\pi}\int_{0}^{2\pi}\int_{0}^{\pi}\boldsymbol{P}_{\chi\phi\theta}\sin(\theta)d\chi d\phi d\theta
\end{equation}
where $\chi, \phi, \theta$ are the Euler angles and $\boldsymbol{P}_{\chi\phi\theta}$ is the induced polarization in the laboratory frame for a given molecular orientation.
The integral over $\phi$, $\theta$ was approximated using the Lebedev quadrature \cite{Lebedev1999AQF} of order 7, which gives 26 points on a sphere ($\phi$, $\theta$).
Integration over $\chi$ was performed using the trapezoid method including 8 rotations around each Lebedev axis. 
The radiation emitted by the medium is $\boldsymbol{E}(\omega) = E_x(\omega) \boldsymbol{e}_x + E_y(\omega) \boldsymbol{e}_y$, where $E_x$ and $E_y$ are the achiral and chiral polarisation,
\begin{align}
E_x(\omega) &\propto \omega^2 \boldsymbol{P}(\omega) \cdot \hat{\boldsymbol{e}}_x, \\
E_y(\omega) &\propto \omega^2 \boldsymbol{P}(\omega) \cdot \hat{\boldsymbol{e}}_y.
\end{align}
Note that the $z$ component of the induced polarisation is parallel to the propagation direction of the optical field and thus it cannot produce a macroscopic response.
The ellipticity ($\varepsilon$) and rotation angle ($\gamma$) of the emitted radiation are given by
\begin{align}
\gamma &= 0.5 \arctan\left(\frac{2|E_{x}||E_{y}|\cos(\delta)}{|E_{x}|^2-|E_{y}|^2}\right), \\
\varepsilon &=\frac{1-\sqrt{1-\alpha^2}}{\alpha},
\end{align}
where $\delta = \arg(E_y) - \arg(E_y)$ and $\alpha = \frac{2\tan(\psi)\sin(\delta)}{1+\tan(\psi)^2}$, with $\tan(\psi) = \frac{|E_{y}|}{|E_{x}|}$.

\section*{Supplementary Information}

\begin{figure}
\includegraphics[width=0.48\textwidth]{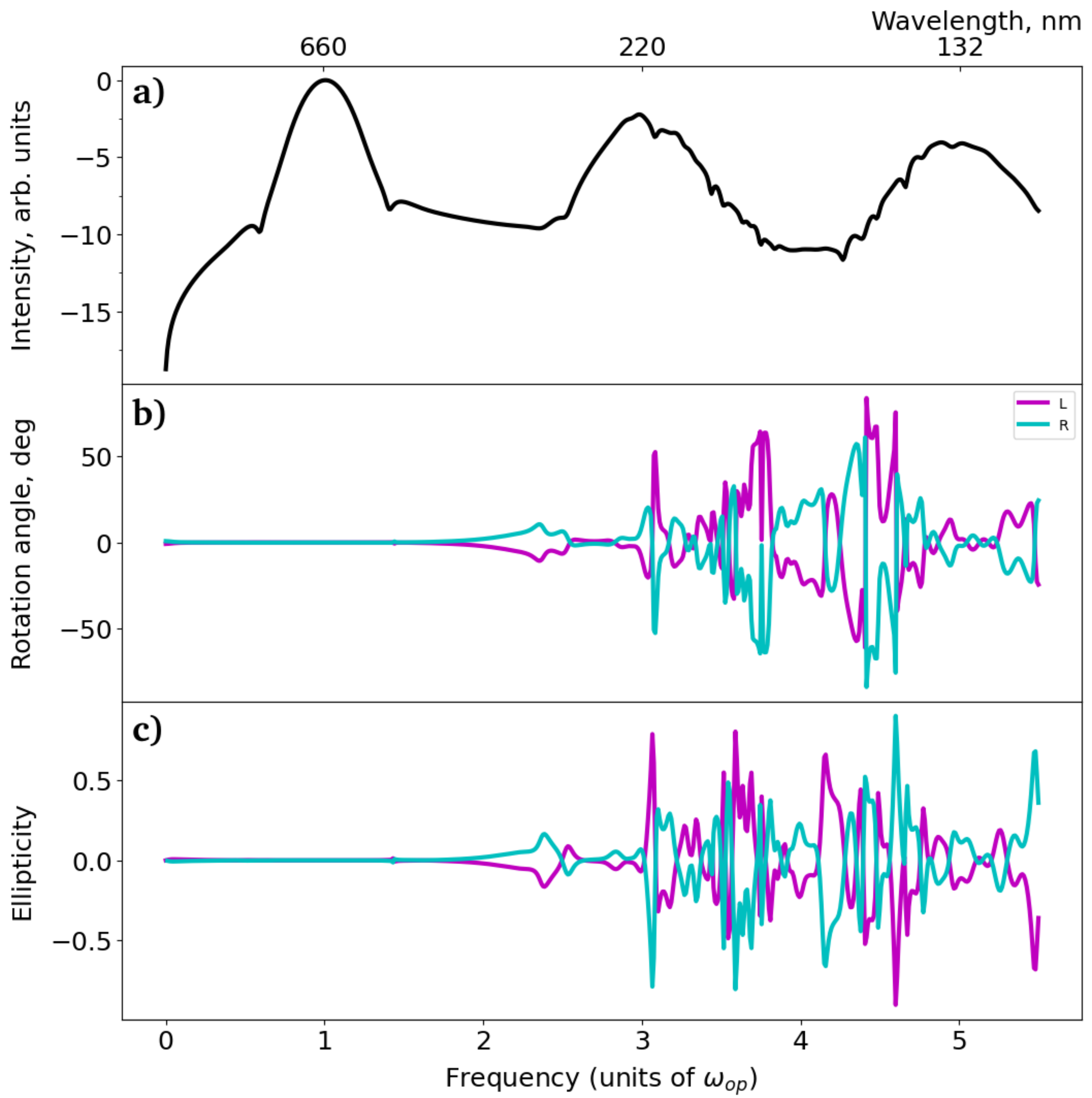}
\caption{\label{fig_results_extended}
\textbf{Non-linear response of randomly oriented carvone molecules ($\boldsymbol{\mathrm{CEP} = 0 }$)} driven by the setup of Fig. \ref{fig_picture}.
\textbf{a} Intensity of the harmonic emission in the range from 0 to $5.5\omega_{\mathrm{op}}$ as a function of the output frequency (lower axis) and wavelength (upper axis).
The strength of the harmonic emission is identical in opposite enantiomers.
\textbf{b,c,} Enantiosensitive polarisation of the radiation emitted from the L (magenta) and R (cyan) enantiomers: rotation angle of the major component of the polarisation ellipse (\textbf{b}) and ellipticity (\textbf{c}).
}
\end{figure}

\bibliography{bibliograpy}
\end{document}